\documentclass[prd,onecolumn,nofootinbib,aps,tightenlines,preprintnumbers]{revtex4-1}

\usepackage{color}
\usepackage{amssymb}
\usepackage{amsmath}
\usepackage[shortlabels]{enumitem}
\usepackage{rotating}
\usepackage{graphicx}
\usepackage[colorlinks=true,citecolor=blue,urlcolor=blue]{hyperref}

\newcommand{\cm}{\ensuremath{\mathrm{cm}}}
\newcommand{\s}{\ensuremath{\mathrm{s}}}
\newcommand{\GeV}{\ensuremath{\mathrm{GeV}}}

\newcommand{\PhiPP}{\Phi_\textrm{PP}}
\newcommand{\Aeff}{A_\textrm{eff}}
\newcommand{\Abareff}{\bar{A}_\textrm{eff}}
\newcommand{\Nbar}{\overline{N}}

\newcommand{\bgd}{\textrm{bgd}}
\newcommand{\sig}{\textrm{sig}}
\newcommand{\obs}{\textrm{obs}}
\newcommand{\bound}{\textrm{bound}}
\newcommand{\Emin}{E_\textrm{min}}
\newcommand{\Emax}{E_\textrm{max}}

\begin{document}

\preprint{\hbox{PREPRINT UH511-1306-2019}}

\title{\texttt{MADHAT}: Model-Agnostic Dark Halo Analysis Tool}
\author{Kimberly K. Boddy}
\affiliation{\mbox{Department of Physics \& Astronomy,
Johns Hopkins University, Baltimore, MD 21218, USA}}
\author{Stephen Hill}
\affiliation{Department of Physics and Astronomy,
University of Hawai'i, Honolulu, HI 96822, USA}
\author{Jason Kumar}
\affiliation{Department of Physics and Astronomy,
University of Hawai'i, Honolulu, HI 96822, USA}
\author{Pearl Sandick}
\affiliation{\mbox{Department of Physics and Astronomy,
University of Utah, Salt Lake City, UT 84112, USA}}
\author{Barmak Shams Es Haghi}
\affiliation{\mbox{Department of Physics and Astronomy,
University of Utah, Salt Lake City, UT 84112, USA}}

\begin{abstract}
We present the Model-Agnostic Dark Halo Analysis Tool (\texttt{MADHAT}), a numerical tool which implements a Fermi-LAT data-driven, model-independent analysis of gamma-ray emission from dwarf satellite galaxies and dwarf galaxy candidates due to dark matter annihilation, dark matter decay, or other nonstandard or unknown astrophysics.  This tool efficiently provides statistical upper bounds on the number of observed photons in excess of the number expected, based on empirical determinations of foregrounds and backgrounds, using a stacked analysis of any selected set of dwarf targets. It also calculates the resulting bounds on the properties of dark matter under any assumptions the user makes regarding dark sector particle physics or astrophysics. As an application, we determine new bounds on Sommerfeld-enhanced dark matter annihilation in a set of eight dwarfs. \texttt{MADHAT v1.0} includes 58 dwarfs and dwarf candidate targets, and we discuss future planned developments. \texttt{MADHAT} is available and will be maintained at \url{https://github.com/MADHATdm}.
\end{abstract}
\maketitle

\section{Introduction}

Dwarf galaxies (dwarfs) are gravitationally-bound astrophysical objects, notable for having a matter distribution that is heavily dominated by dark matter.
They are an excellent target for indirect dark matter searches: dwarfs contain few baryonic sources of high energy photons that would contaminate a dark matter signal, and prompt photons produced by dark matter annihilation/decay within a dwarf point directly back to the source.
Indeed, searches for photons emanating from dwarfs using Fermi-LAT data provide some of the tightest constraints on dark matter annihilation cross sections (e.g., see Refs.~\cite{Abdo:2010ex,Ackermann:2011wa,Ackermann:2013yva,Ackermann:2015zua,Fermi-LAT:2016uux}).
These published constraints are typically based on specific assumptions about dark matter particle physics and astrophysics and about the dwarfs used in the analysis; for any other choice, one must perform a new and potentially computationally-intensive analysis.
In this paper, we present an efficient numerical tool that allows the user to perform a stacked analysis of a set of dwarf satellite galaxies and dwarf galaxy candidates (hereafter collectively referred to as ``dwarfs'') to determine dark matter annihilation/decay exclusion constraints for any choice of dark matter particle physics or astrophysics model.

We introduce the Model-Agnostic Dark Halo Analysis Tool (\texttt{MADHAT}), an automated numerical implementation of the data-driven formalism introduced in Ref.~\cite{GeringerSameth:2011iw} (a model-agnostic approach is described in Ref.~\cite{Boddy:2018qur}, and other data-driven approaches are described e.g.\ in Refs.~\cite{Mazziotta:2012ux,GeringerSameth:2012sr,Geringer-Sameth:2014qqa,Geringer-Sameth:2015lua,Calore:2018sdx,Geringer-Sameth:2018vjd,Hoof:2018hyn,Linden:2019soa}).
\texttt{MADHAT} incorporates processed Fermi-LAT data for many known dwarfs, allowing for a quick stacked analysis.
The formalism relies on statistical estimates of the foreground/background gamma-ray flux along the observational line of sight.
Examining the gamma-ray flux slightly off-axis from a dwarf target provides an estimate for the number of photons attributable to astrophysical foreground/background expected to arrive on-axis from the target region.
The user may choose the set of dwarf targets to stack, and the total number of photons observed from the directions of the targets places a statistical bound on the expected number of photons attributable to prompt dark matter annihilation/decay or, potentially, other non-standard or unknown astrophysics.
This bound is conservative, since other possible anomalous sources of photons are not considered here.
For example, unresolved point sources could be located near the line of sight to one or more targets and could therefore contribute to an excess of photons, thereby weakening the constraint on dark matter annihilation.  
Finally, the user can specify the particle properties of dark matter and the annihilation ($J$) and decay ($D$) factors of the dwarf targets to determine a statistical bound on the dark matter annihilation cross section or decay rate, respectively; the user can also take the \texttt{MADHAT} output and use it to constrain more exotic dark matter models.

The bounds obtained from this formalism are not necessarily the strongest; a dedicated analysis designed to study a particular particle physics or astrophysics scenario would likely generate stronger constraints, though perhaps not dramatically so~\cite{Boddy:2018qur}.
The advantage of this formalism is that it can easily be applied to any particle physics or dark matter halo model, as those assumptions only enter at the final stage of the analysis.
This model independence is especially useful if \texttt{MADHAT} is incorporated into a larger framework to produce global constraints on dark matter, for which computational speed and model flexibility are crucial.
Moreover, as new dwarfs are discovered and as the Fermi-LAT acquires more publicly-available data, \texttt{MADHAT} will be updated with new files of processed Fermi-LAT data.

The plan of this paper is as follows.
In Sec.~\ref{sec:framework}, we review the framework of the analysis.
In Sec.~\ref{sec:implementation} we describe how to use the code that implements this analysis.
As an application, we use this code to obtain new bounds on Sommerfeld-enhanced dark matter annihilation in Sec.~\ref{sec:example}.
We conclude with a discussion in Sec.~\ref{sec:conclusions}.

\section{Analysis Framework}
\label{sec:framework}

\texttt{MADHAT} allows users to perform a stacked analysis of dwarfs with Fermi-LAT data to place bounds on the number of photons not attributable to foregrounds/backgrounds, and, in particular, on dark matter annihilation.
The methodology follows that in Refs.~\cite{GeringerSameth:2011iw,Boddy:2018qur}, in which the Fermi-LAT data, assumptions about the dwarf halo properties, and assumptions about dark matter particle model can be treated as modular components of the analysis.
Using this framework gives the analysis flexibility, allowing \texttt{MADHAT} to be a versatile tool that is applicable over a wide range of dark matter scenarios.

We summarize the analysis pipeline here and refer the reader to Refs.~\cite{GeringerSameth:2011iw,Boddy:2018qur} for more detail.
In Sec.~\ref{sec:framework-data}, we briefly discuss how we use Fermi-LAT data to obtain necessary gamma-ray information associated with each dwarf, which is treated as input to \texttt{MADHAT}.
We then describe how \texttt{MADHAT} uses this data to set limits on the number of excess gamma rays arising from anomalous sources in the directions of the dwarfs in Sec.~\ref{sec:framework-Nobs} and subsequently set limits on dark matter properties in Sec.~\ref{sec:framework-PhiPP}.

\subsection{Processing Fermi-LAT data}
\label{sec:framework-data}

The stacked analysis implemented in \texttt{MADHAT} relies on the observed number of photon counts from each dwarf in question and from its surrounding region.
There is no spectral analysis in our framework, so all photons in an energy range $\Emin < E_\gamma < \Emax$ are incorporated.
While we are concerned only with photon counts in Sec.~\ref{sec:framework-Nobs}, we need the corresponding flux in Sec.~\ref{sec:framework-PhiPP} to connect to dark matter properties.
The flux from a given target is the number of photons divided by the effective area $\Aeff$ of the Fermi-LAT and the observation time.
In general, the $\Aeff$ varies with the gamma-ray energy; however, in the energy range between $1~\GeV$ and $100~\GeV$, it is fairly insensitive to the gamma-ray energy and may be treated as constant.
Therefore, we choose $\Emin = 1~\GeV$ and $\Emax = 100~\GeV$ and set the effective area to its average value $\Abareff$.
Doing so allows us to handle the data separately from the spectral information of the background/foreground and the dark matter annihilation/decay.

Processing Fermi-LAT data can be nontrivial, so we extract the necessary gamma-ray information needed for the analysis to distribute with \texttt{MADHAT}.
We include the following for each dwarf target:
\begin{itemize}
\item{the effective exposure $\Abareff T_\obs$, where $T_\obs$ is the total observation time of the target,}
\item{the number of observed photons $N_\obs$ with $\Emin < E_\gamma < \Emax$ arriving from the target region, a cone with an opening angle of $1^\circ$ centered at the target, and}
\item{the probability mass function (PMF) for foreground/background photons with $\Emin < E_\gamma < \Emax$ within $1^\circ$ sample regions.
  As described in Ref.~\cite{Boddy:2018qur}, the PMF is a histogram of the number of photons observed in $10^5$ randomly chosen sample regions within $10^\circ$ of the target, excluding any sample regions that overlap with the target region or are centered within $1.3^\circ$ of a known point source.}
\end{itemize}
Increasing the size of the masked regions has negligible effects on our overall results.
We consider a total of 58 dwarfs, listed in Tables~\ref{Tab:DsphParametersA} and \ref{Tab:DsphParametersB}, using their location information from the \texttt{dmsky}\footnote{\url{https://github.com/fermiPy/dmsky}} tool.
To exclude known point sources, we obtain their locations from the Fermi 3FGL catalog~\cite{Acero:2015hja}.

We utilize Fermi-LAT Pass 8R3 data~\cite{Atwood:2009ez,Bruel:2018lac} in the mission elapsed time range of 239557417 to 585481831 seconds (11 years), using the Fermi Science Tools \texttt{1.0.10}\footnote{\url{https://github.com/fermi-lat/Fermitools-conda}} and \texttt{FermiPy}\footnote{\url{https://github.com/fermiPy/fermipy}}.
We apply the selection specifications \texttt{evclass=128}, \texttt{evtype=3}, and \texttt{\protect{\detokenize{zmax=100}}} and the filter `\texttt{\protect{\detokenize{(DATA_QUAL>0)&&(LAT_CONFIG==1)}}}'.
Compared to our previous work~\cite{Boddy:2018qur}, we consider more dwarfs and include more Fermi-LAT data.
We obtain the effective area and time from the exposure map, generated by the Fermi Science Tools \texttt{1.0.10}. We use \texttt{P8R3\_SOURCE\_V2}\footnote{\url{https://fermi.gsfc.nasa.gov/ssc/data/access/lat/BackgroundModels.html}} for the instrument response functions (IRF).

\subsection{Constraining excess events}
\label{sec:framework-Nobs}

With the PMF and observed photon count for each dwarf, a straightforward analysis to perform is constraining the excess number of photons over background/foreground that arise from an anomalous source in the target region.
Since we are ultimately interested in constraining photons from dark matter annihilation/decay, we gain statistical power by stacking all dwarfs, all of which would have annihilation/decay signals.
We note, however, that the user can run \texttt{MADHAT} for a single dwarf target to assess the probability of excess photons arising from, for instance, a previously unknown point source.
In any case, we assume that the total number of photon counts from the anomalous sources in all target regions follows a single Poisson distribution.

The user can define a subset of the 58 dwarfs included in \texttt{MADHAT} to use for the stacked analysis, in which we imagine overlaying the photon-count maps for those dwarfs.
\texttt{MADHAT} convolves the PMFs for all the selected targets to obtain the joint PMF for foreground/background photons arriving from within an arbitrary $1^\circ$ sample region in the vicinity of the stacked target region.
It then convolves the joint PMF with the signal Poisson distribution, given some expected mean number of signal counts $\Nbar_\sig$, to obtain the total probability distribution.
The code adjusts $\Nbar_\sig$ until the probability of observing at least the total number of observed photons $N_\obs$ arising from the stacked target region matches a user-specified value $0 < \beta < 1$; we denote this count number as $N_\bound (\beta)$.
For further details on this computation, we refer the reader to Ref.~\cite{Boddy:2018qur}.
Setting $\Nbar_\sig > N_\bound (\beta)$ shifts the total probability distribution to the right towards higher number counts, leaving the observed $N_\obs$ lying further in the left tail of the distribution; thus, observing $N_\obs$ becomes more disfavored for this total distribution.
In this sense, we say that $N_\bound (\beta)$ represents an upper limit on $\Nbar_\sig$ with a confidence level (C.L.) of $\beta$.
We urge caution in interpreting $\beta$ as a C.L., particularly at small $N_\bound (\beta)$, for which there is little difference between the total distribution and the PMF, or at low $\beta$.

At this point, we emphasize that the choice of targets should be independent of the Fermi-LAT data itself.
Using a subset of targets that all have slight excesses above the mean of their PMF distributions would bias the analysis and degrade the interpretation of $\beta$.
It is the responsibility of the user to ensure that their dwarf sets are constructed in an unbiased manner.

The analysis thus far has assumed only that there is an anomalous source of gamma rays originating from the dwarf target region that is not present in the nearby surrounding area.
We now attribute any possible excess of photons to dark matter annihilation/decay.
As previously mentioned, it may be that a previously unknown point source near a target region could be contributing to the observed number of photons.
Disregarding such possibilities and assuming all anomalous events are due to dark matter, we obtain a conservative limit on $N_\bound$.

\subsection{Constraining dark matter properties}
\label{sec:framework-PhiPP}

We may now translate $N_\bound(\beta)$ into a bound on specific dark matter properties.
In particular, we focus on signals from dark matter annihilation, since dwarfs are ideal systems to search for such processes.
Our analysis holds equally well for decay, and we note the differences throughout this section.

For a broad class of dark matter scenarios, the photon flux arising from dark matter annihilation can be factorized into a particle physics factor $\PhiPP$ and an astrophysical $J$-factor (referred to as the $D$-factor for decay).
The factor $\PhiPP$ is independent of the target, and the $J$-factor is determined by properties of the dark matter halo (it may also depend on the dark matter microphysics if the annihilation cross section is velocity-dependent~\cite{Robertson:2009bh,Ferrer:2013cla,Boddy:2017vpe,Bergstrom:2017ptx,Zhao:2017dln,Petac:2018gue,Lacroix:2018qqh,Boddy:2018ike,Boddy:2019wfg,Boddy:2019qak}). 
The number of expected signal photons due to dark matter annihilation/decay in the $i$th target can thus be written as
\begin{equation}
  \Nbar_S^i = \PhiPP \times J^i (\Delta \Omega) \times (\Abareff^i T_\obs^i) \ ,
\end{equation}
where $J^i (\Delta \Omega)$ is the $J$-factor of the $i$th target integrated over the solid angle $\Delta \Omega$ and $(\Abareff^i T_\obs^i)$ is the exposure of the $i$th target.
It should be emphasized, this decomposition is an approximation.
But it is a good approximation for our purposes and a necessary one in order to obtain model-independent limits.
A more exact approach would be to convolve the angular extent of the emission with the point spread function (PSF), but this would require a choice of dark matter profile (NFW, Burkert, etc.).
Moreover, the PSF is strongly energy-dependent within the range $1-100~\GeV$, so one would have to include in this convolution the photon spectrum arising from dark matter annihilation, thus requiring a choice of particle physics model.
Given the typical systematic uncertainties in the $J$-factors, this approximation is not a major source of uncertainty.
At $1~\GeV$, the PSF is about $0.8^\circ$; given the size of the sample region ($1^\circ$), and assuming a dSph with an angular size of $0.2^\circ$, it can lead to an $\mathcal{O}(1)$ suppression in the photon count, yielding a conservative bound on dark matter annihilation.
But note that, at higher energies ($\sim 10-100~\GeV$), the PSF decreases rapidly, and the effect of the PSF on the photon count in the sample region becomes much smaller.
Although many standard choices of the annihilation final state (e.g., $b\bar{b}$) produce photon spectra that peak at lower energies, there are also non-standard choices (e.g., cascade decays such as $\chi\chi \rightarrow \phi\phi \rightarrow 4\gamma$) that can produce spectra peaked at higher energies.
Our approximation allows one to obtain constraints which can be applied to non-standard choices as well as the more standard particle and astrophysics models.

Consider, for example, the case in which dark matter consists of a single self-conjugate particle species with an annihilation cross section $\sigma_A v = (\sigma_A v)_0 \times S(v)$, where $(\sigma_A v)_0$ is a constant, independent of the relative velocity $v$.
We then have
\begin{align}
  \PhiPP &= \frac{(\sigma_A v)_0}{8\pi m_X^2}
  \int_{\Emin}^{\Emax} dE_\gamma \frac{dN_\gamma}{ dE_\gamma} , \nonumber\\
  J(\Delta\Omega) &= \int_{\Delta\Omega} d\Omega \int d\ell \int d^3 v_1 \, d^3 v_2 \,
  f\left(\boldsymbol{r}, \boldsymbol{v}_1\right)
  f\left(\boldsymbol{r}, \boldsymbol{v}_2\right) S(|\boldsymbol{v}_1-\boldsymbol{v}_2|) \ ,
  \label{eqn:StandardAssumptions}
\end{align}
where $m_X$ is the dark matter mass, $dN_\gamma / dE_\gamma$ is the photon spectrum per annihilation, $\ell$ is the distance along the line of sight, and $f(\boldsymbol{r},\boldsymbol{v})$ is the dark matter velocity distribution.
We do not convolve the photon spectrum with a smearing function to account for the Fermi-LAT energy resolution; since we are integrating over a relatively large energy range, the effect on our result is negligible.
In the most commonly studied case of $s$-wave annihilation, the cross section is velocity-independent such that $S(v)=1$ and the $J$-factor reduces to $J = \int_{\Delta \Omega} d\Omega \, d\ell \, \rho^2$, where $\rho$ is the dark matter density profile.
Note that if the dark matter particle and antiparticle were distinct (with equal abundances), then the expression for $\PhiPP$ in Eq.~\eqref{eqn:StandardAssumptions} would be multiplied by an extra factor of $1/2$.
For dark matter decay with a decay rate $\Gamma$, we would substitute $(\sigma_A v)_0/2m_X \to \Gamma$ and $J \to \int_{\Delta\Omega} d\Omega \int d\ell \, \rho$.

We can obtain the bound $\PhiPP^\bound (\beta)$ at the $\beta$ C.L. from setting $N_{\bound}(\beta) = \sum_i N_S^i$ and dividing by $\sum_i J^i (\Delta\Omega) \times (\Abareff^i T_\obs^i)$.
The exposures are provided in \texttt{MADHAT} as part of the data processing explained in Sec.~\ref{sec:framework-data}, but $J$-factors for the chosen subset of dwarfs are also needed.
There are many methods for estimating $J$-factors from stellar data, and results can vary widely for different underlying assumptions.
For example, the $J$-factors can change significantly if one does not assume a spherically-symmetric dark matter distribution or if one assumes velocity-dependent dark matter annihilation, in which case the $J$-factor depends on the dark matter velocity distribution, as in Eq.~\eqref{eqn:StandardAssumptions}.
\texttt{MADHAT} is distributed with example input files of $J$-factors (and their uncertainties) from various sources, discussed more in the following section.
The user may also create their own input files with their choice of $J$-factors.

Finally, we can translate the bound on $\PhiPP$ to a bound on parameters of a specific dark matter model.
In particular, \texttt{MADHAT} assumes $\PhiPP$ is given by Eq.~\eqref{eqn:StandardAssumptions}.
The user provides the values of $m_X$ and the integrated energy spectrum per annihilation, and \texttt{MADHAT} produces a bound on $(\sigma_A v)_0$ at the $\beta$ C.L.
Performing this procedure over a range of $m_X$ results in an exclusion curve at the $\beta$ C.L. in the $(m_X, (\sigma_A v)_0)$-plane.
For a more generic scenario, the user must convert $\PhiPP^\bound (\beta)$ themselves to constrain the parameters of their particular dark matter model.

At this point, it is worthwhile to reiterate the assumptions underlying this analysis.
In particular, it is assumed that the number of $1-100~\GeV$ photons arriving from within a $1^\circ$ cone centered at a dwarf galaxy can be drawn from a statistical distribution consisting of two pieces.
The first piece is an astrophysical foreground/background distribution, obtained from the histogram of the number of photons arriving from many such cones located near the dwarf galaxy, but slightly off axis.
The second piece is an additional Poisson-distributed source of photons arriving from along the line-of-sight to the dwarf galaxy.
Given the astrophysical foreground/background distributions and the number of actual photons observed to arrive from any set of dwarfs (both obtained from Fermi data), the result of the analysis is an upper bound on the expected number of photons yielded by the Poisson-distributed source ($N_\bound$).

A priori, no assumption is made regarding the nature or origin of the Poisson-distributed source of photons; thus far, the analysis is entirely data-driven, with no assumptions beyond the forms of the statistical distributions.
But one may add the additional assumption that the Poisson-distributed source is dark matter annihilation within the dwarfs, with dwarf galaxy $J$-factors which are specified by the user.
If this assumption is made, then the bound on the expected number of photons is translated into a bound on the factor $\PhiPP$, which encodes all of the dark matter particle physics information.
In performing this translation, uncertainties in the $J$-factor are treated as purely systematic.
A statistical bound on $\PhiPP$ is obtained by assuming a particular choice of the $J$-factors, and if the J-factors are varied within their systematic uncertainties, the statistical bound on $\PhiPP$ varies as well.
We do not assume that the true $J$-factors are drawn from any particular distribution.
But as a rough estimate of how statistical bound on $\PhiPP$ can vary with the uncertainties in the $J$-factor, we present systematic uncertainties in bounds on $\PhiPP$ which are obtained by either varying all $J$-factors upward or varying all $J$-factors downward by their systematic uncertainties.
Finally, if it is assumed that dark matter consists entirely of a single species of real particle, with a known annihilation photon spectrum, then the statistical bound on  $\PhiPP$ is translated into a bound on the annihilation cross section.

Note that the Fermi-LAT has a non-trivial PSF.
Although it can be as large as $\sim 0.8^\circ$ at $\sim 1~\GeV$, it is strongly energy-dependent.
As such, it cannot be corrected for without enforcing particular assumptions about the dark matter distribution and about the photon spectrum arising from dark matter annihilation.
In order to remain agnostic about the dark matter particle and astrophysics, we do not attempt to correct for the PSF.
This does not affect the PMF significantly, but can suppress the count of photons arriving from the signal region, since the tail of the PSF may lie outside the signal region.
But the uncertainty produced by our treatment of the PSF is relatively small compared to the typical systematic uncertainties in the $J$-factor.

\section{Implementation}
\label{sec:implementation}

\texttt{MADHAT} is available at \url{https://github.com/MADHATdm}.
The \texttt{MADHAT} package includes code written in C++, processed Fermi-LAT data for 58 dwarfs, ten pre-defined dwarf sets, four dark matter model files for annihilation to two-body final states, and templates for defining additional dwarf sets and dark matter models.
Detailed instructions for installation and execution can be found on the \texttt{MADHAT} wiki.\footnote{\url{https://github.com/MADHATdm/MADHAT/wiki}}
In order to run \texttt{MADHAT}, the user must specify, at a minimum, the set of dwarfs to analyze and the $\beta$ C.L. for the limits.
The user may also specify the dark matter mass and integrated photon spectrum between $1~\GeV$ and $100~\GeV$, which are used to calculate limits on $(\sigma_A v)_0$.  Input formatting and argument specification are described briefly below.

There are three options for running \texttt{MADHAT}:
\begin{enumerate}[A)]
\item Specify the set of dwarfs to be analyzed, the confidence level ($\beta$) for $N_\bound$, and the dark matter model parameters.
  This option requires three arguments to run:
  \begin{verbatim} ./madhat [dwarfset.dat] [beta] [model.in] \end{verbatim}
  \texttt{MADHAT} will read \texttt{[model.in]} and calculate output for each line until it reaches the end of the \texttt{[model.in]} file.
  Output will print to a file in the \texttt{Output} directory named \texttt{[model\_dwarfset\_beta.out]} with the following columns:
  mass, integrated photon spectrum, $\beta$, $N_\bound (\beta)$, $\PhiPP$, +$d\PhiPP$, --$d\PhiPP$, $(\sigma_A v)_0$, +$d(\sigma_A v)_0$, --$d(\sigma_A v)_0$.
  Note that $\pm d\PhiPP$ and $\pm d(\sigma_A v)_0$ are the uncertainties in $\PhiPP$ and $(\sigma_A v)_0$ obtained by varying all $J$-factors up or down by one standard deviation.
\item Specify the set of dwarfs to be analyzed, the confidence level ($\beta$) for $N_\bound$, and the mass and integrated photon spectrum for single dark matter model point:
  \begin{verbatim} ./madhat [dwarfset.dat] [beta] [mass] [integrated spectrum] \end{verbatim}
  This option is a duplicate of option A, but allows the user to quickly check a single model point.
  Output will print to the screen.
\item Specify the set of dwarfs to be analyzed and the confidence level ($\beta$) for $N_\bound$.
  This option requires two arguments to run:
  \begin{verbatim} ./madhat [dwarfset.dat] [beta] \end{verbatim}
  Output will print to screen in the following format: $\beta$, $N_\bound (\beta)$, $\PhiPP$, +$d\PhiPP$, --$d\PhiPP$.
\end{enumerate}

Arguments to the \texttt{MADHAT} executable must be specified as follows:
\begin{itemize}
\item \texttt{[dwarfset.dat]} is a file containing the parameters for the dwarfs the user would like to analyze and must be located in the directory \texttt{Input}.  This file must contain, at a minimum, the ID numbers for the dwarfs to be considered.  It may also contain $J$-factors for each dwarf and $J$-factor errors.  Note that errors are not necessary, and if omitted, errors on $\pm d\PhiPP$ will print as zeros. Similarly, if the file only contains dwarf ID numbers, $\PhiPP$ will also print as zero. There are ten \texttt{[dwarfset.dat]} files included in the \texttt{MADHAT} package, plus a template file, \texttt{SetTemplate.dat}.
\item \texttt{[beta]} is a number between zero and one that specifies the confidence level (e.g., 0.95 for 95\% C.L.).
\item \texttt{[model.in]} is a file containing a list of dark matter masses and integrated photon spectra, each of which must be floating point numbers, and must be located in the directory \texttt{Input}.  There are four files of this type included in the \texttt{MADHAT} package: \texttt{DMbb.in}, \texttt{DMWW.in}, \texttt{DMmumu.in}, and \texttt{DMtautau.in}, which are basic model files for dark matter annihilation to $b\bar{b}$, $W^+W^-$, $\mu^+\mu^-$, and $\tau^+\tau^-$, respectively, tabulated using the PPPC4DMID spectra~\cite{Cirelli:2010xx, Ciafaloni:2010ti}.  We also include a template/test file, \texttt{dmtest.in}.
\item The dark matter mass and integrated photon spectrum for a single model point may also be specified directly as arguments to the executable (run option B).
  In this case, both \texttt{[mass]} and \texttt{[integrated spectrum]} must be positive floating point numbers.
\end{itemize}
For details on formatting of the input files, please consult the wiki.

In Tables~\ref{Tab:DsphParametersA} and~\ref{Tab:DsphParametersB}, we list the 58 dwarfs 
(and dwarf candidates) included in the \texttt{MADHAT} package, along with the Fermi-LAT exposure ($\Abareff T_\obs$), the average of number of photons from the PMF ($\Nbar_\bgd$), and the number of photons observed from the direction of the target ($N_\obs$).
\texttt{MADHAT} contains several predefined sets of dwarf targets with $J$-factors and uncertainties found by previous studies.
We include the sets defined in our previous work~\cite{Boddy:2018qur}: {\it Set 1} (and subsets {\it 1a, 1b, 1c})\footnote{The $J$-factors used in this set were obtained from Ref.~\cite{Geringer-Sameth:2014yza}, with the exception of Reticulum II, which was obtained from Ref.~\cite{Simon:2015fdw}. These were the values used in Ref.~\cite{Fermi-LAT:2016uux}.}, {\it 2}~\cite{Evans:2016xwx}, {\it 3}~\cite{Hayashi:2016kcy}, {\it 4}~\cite{Ichikawa:2016nbi,Ichikawa:2017rph}, and {\it 5}~\cite{Boddy:2017vpe}.
We introduce {\it Set 6} with $J$-factors from Ref.~\cite{Pace:2018tin}; we take the $J$-factors integrated over the largest angular cone of $1^\circ$ (to encompass as much of the dwarf as possible) and take the values for Horologium I, Reticulum II, and Tucana II that used the parameters from Ref.~\cite{Koposov:2015cua}.
We also include {\it Set 7} with $J$-factors for Sommerfeld-enhanced annihilation in the Coulomb limit for 8 dwarfs from Ref.~\cite{Petac:2018gue}; we take the $J$-factors determined for an Navarro-Frenk-White (NFW) profile, using the Eddington method.
For each of the 58 targets used by \texttt{MADHAT}, Tables~\ref{Tab:DsphParametersA} and \ref{Tab:DsphParametersB} indicate which sets the targets belong to and the value of the $J$-factor used in each set.
Note that {\it Sets 1a,1b,1c} are all subsets of {\it Set 1} with no change in the values of the $J$-factors, and membership in these subsets is indicated in Tables~\ref{Tab:DsphParametersA} and~\ref{Tab:DsphParametersB} by a check mark.
The user can choose to analyze a different set of dwarfs with a different choice of $J$-factors by simply creating a new \texttt{[dwarfset.dat]} file.
Filenames with the format \texttt{SetN.dat}, where N is an integer, are reserved for dwarf sets defined in future versions of \texttt{MADHAT}.

\section{An Application}
\label{sec:example}

We illustrate the utility of this tool by using it to perform a new analysis.
We consider a scenario in which dark matter annihilation is Sommerfeld-enhanced in the Coulomb limit.
In this case, the dark matter annihilation cross section is enhanced at small relative velocities; in particular, $\sigma_A v = (\sigma_A v)_0 \times S(v)$, where $S(v) = (v/c)^{-1}$~\cite{ArkaniHamed:2008qn,Feng:2010zp}.
The standard results for $s$-wave $J$-factors cannot be used; instead, an effective $J$-factor must be derived from the dark matter velocity distribution using Eq.~\eqref{eqn:StandardAssumptions}.

\begin{figure}[t]
  \centering
  \includegraphics[width=0.49\textwidth]{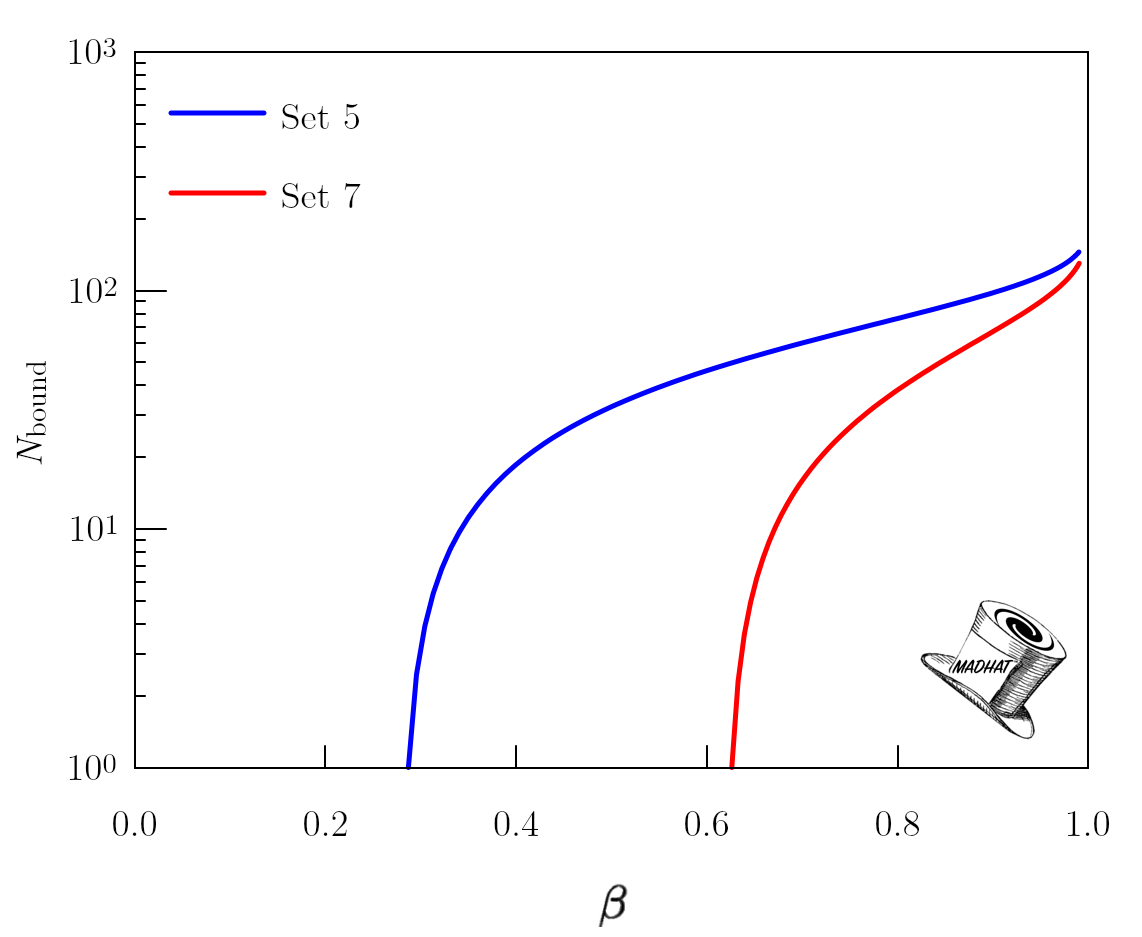}
  \includegraphics[width=0.49\textwidth]{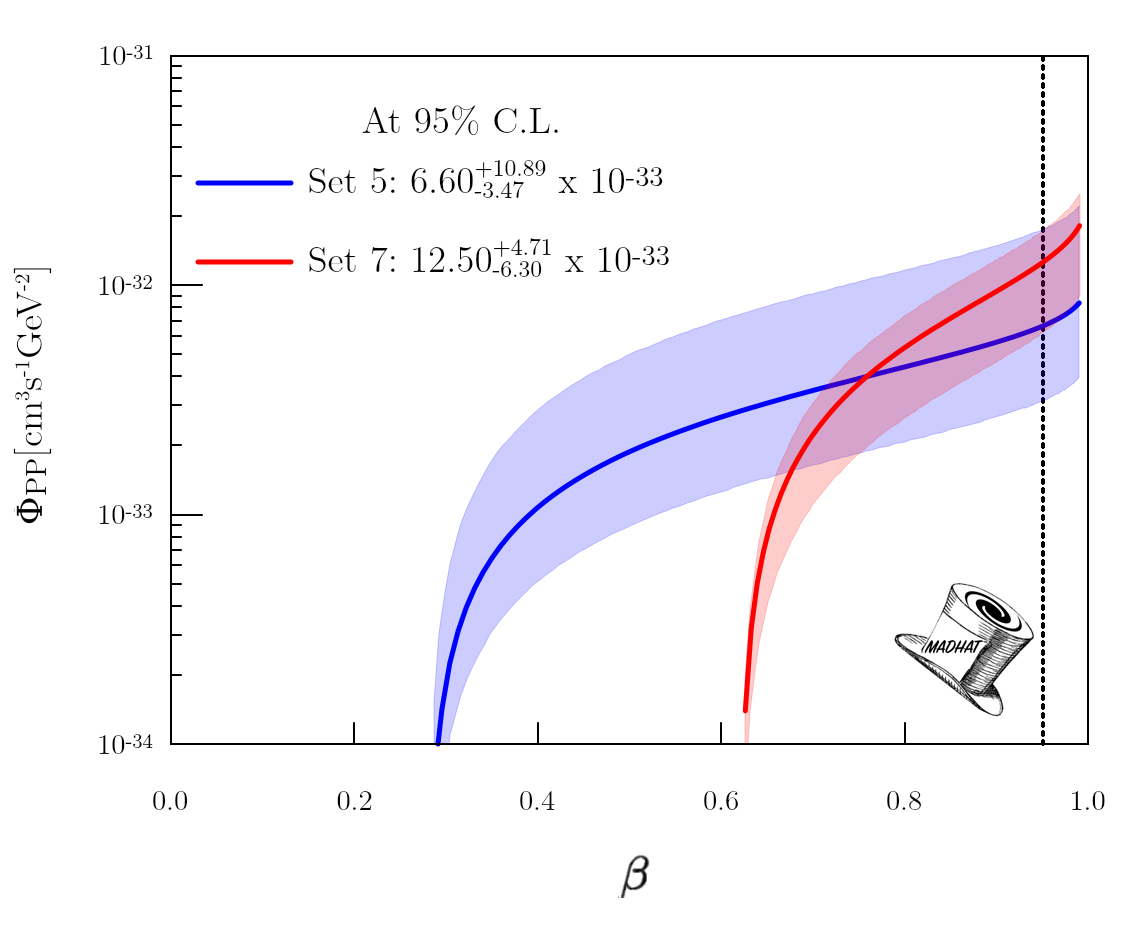}
  \caption{$N_\bound (\beta)$ (left), and $\PhiPP^\bound (\beta)$ (right) as functions of $\beta$, for {\it Set 5} (blue)~\cite{Boddy:2017vpe} and {\it Set 7} (red)~\cite{Petac:2018gue}.
    The uncertainty  band for $\PhiPP^\bound (\beta)$ is determined by varying all of the effective $J$-factors either up or down by through their 1$\sigma$ uncertainties.}
  \label{fig:N_PhiPPbound}
\end{figure}

There have been a variety of calculations of dwarf effective $J$-factors for the case of Sommerfeld-enhanced dark matter annihilation~\cite{Boddy:2017vpe,Bergstrom:2017ptx,Petac:2018gue}, but because there is no unambiguous way to reconstruct the dark matter velocity distribution from stellar data, the approaches of these papers have varied widely.
We use \texttt{MADHAT} to perform an updated analysis from Ref.~\cite{Boddy:2018qur} for the 5 targets in {\it Set 5} and a new analysis for the 8 targets in {\it Set 7}.
The $J$-factors for both of these sets are effective $J$-factors for Sommerfeld-enhanced
dark matter annihilation in the Coulomb limit (with the  dark matter self-coupling taken to be $\alpha_X = 0.01$), though of the 8 targets in {\it Set 7}, only two are in common with the 5 targets in {\it Set 5}~\cite{Boddy:2017vpe}.
Moreover, the approaches to determine the effective $J$-factors in Ref.~\cite{Petac:2018gue} and Ref.~\cite{Boddy:2017vpe} are different.
Although both works assumed a pure NFW profile and used the Eddington inversion formula to obtain the velocity distribution from the density profile, they used different methods to determine the central values and uncertainties in the NFW profile parameters.
In Ref.~\cite{Boddy:2017vpe}, the NFW profile parameters were determined using stellar data and a $V_\textrm{max}-r_\textrm{max}$ relation found in numerical simulations, while in Ref.~\cite{Petac:2018gue}, the parameters were determined using only stellar data.

In Fig.~\ref{fig:N_PhiPPbound}, we present $N_\bound (\beta)$ (left) and $\PhiPP^\bound (\beta)$ (right) for {\it Set 5} (blue) and {\it Set 7} (red).
The uncertainty bands for $\PhiPP^\bound (\beta)$ are determined by varying all of the effective $J$-factors either up or down by the uncertainties given in Ref.~\cite{Boddy:2017vpe} for {\it Set 5} and Ref.~\cite{Petac:2018gue} for {\it Set 7}.
Note that because the included dwarfs, as well as the $J$-factor calculations, differ between the two sets, there are significant differences in $N_\bound (\beta)$ and $\PhiPP^\bound (\beta)$.
The curve for {\it Set 5} begins around $\beta = 0.29$, while the curve for {\it Set 7} does not begin until $\beta = 0.63$.%
\footnote{As previously discussed, interpreting $\beta$ as a C.L.\ is not valid for small $\beta$ or small $N_\bound$ (in the steeply rising portion of the curves in the left panel of Fig.~\ref{fig:N_PhiPPbound}).}
We are able to find, for instance, a limit $N_\bound =46$ or $\PhiPP^\bound = 2.66^{+4.39}_{-1.40} \times 10^{-33}~\cm^3 \s^{-1} \GeV^{-2}$ at the 60\% C.L.\ for {\it Set 5}, while our analysis has no constraining power for {\it Set 7}.

\begin{figure}[t]
  \centering
  \includegraphics[width=0.49\textwidth]{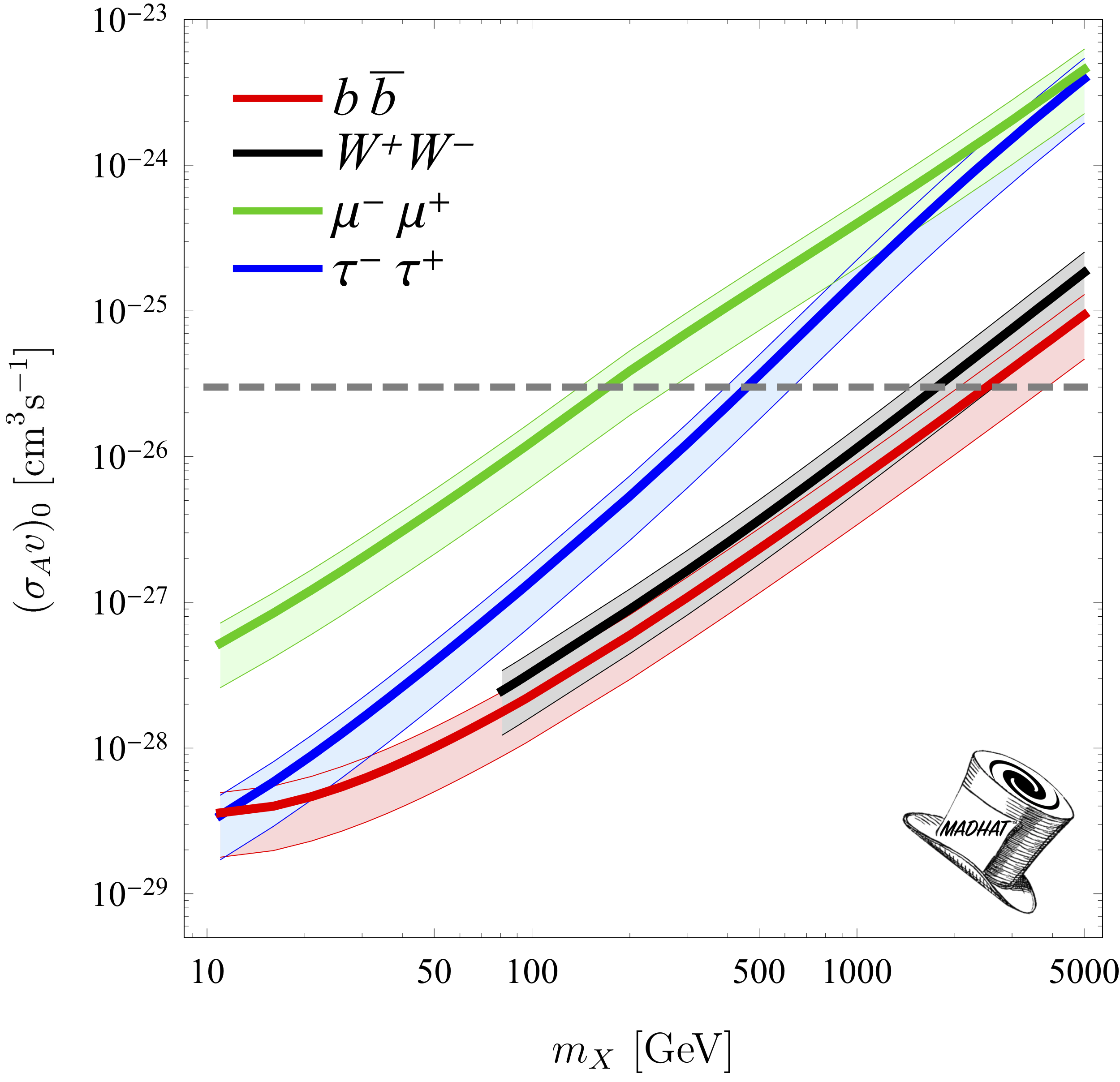}
  \includegraphics[width=0.49\textwidth]{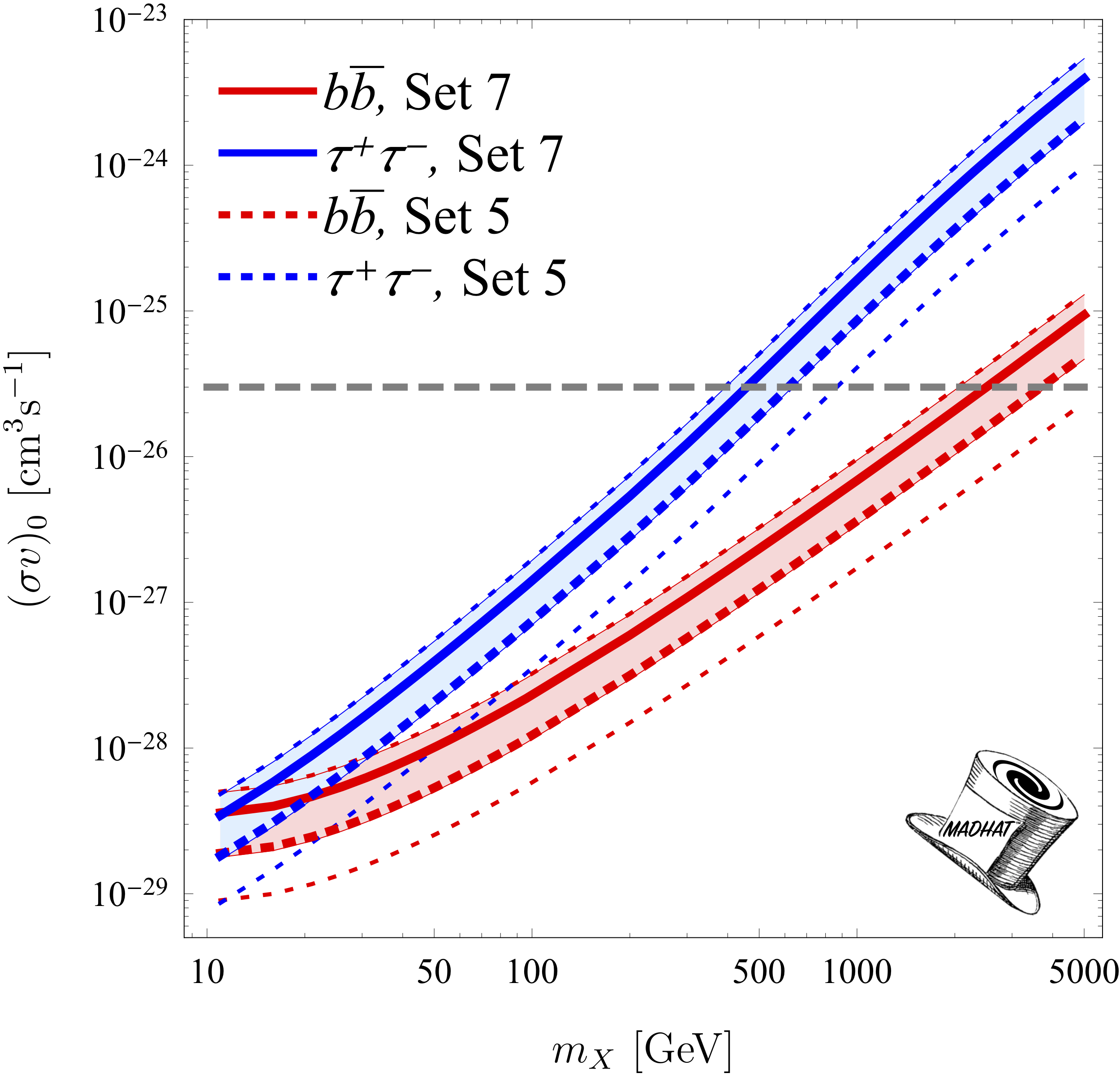}
  \caption{The 95\% C.L. bounds on $(\sigma_A v)_0$ as a function of $m_X$ for dark matter annihilation to standard two-body final states $b\bar{b}$ (red), $W^+ W^-$ (black), $\mu^+\mu^-$ (green), and $\tau^+ \tau^-$ (blue).
  Constraints are presented for Sommerfeld-enhanced annihilations for {\it Set 7} in the left panel.
  The right panel shows a comparison of the constraints from {\it Set 7} (solid contours) and {\it Set 5} (dotted contours) for the $b\bar{b}$ (red) and $\tau^+\tau^-$ (blue) final states.
  In each case, we show the effect on the variation in the 95\% C.L. limits due to the 1$\sigma$ variation in $J$-factors for each model.
  The grey dashed line in each panel indicates a cross section of $(\sigma_A v)_0 = 3 \times 10^{-26}~\cm^3 \s^{-1}$, the thermal annihilation cross section required for a Majorana fermion WIMP dark matter to obtain the observed relic abundance.
  The uncertainty band for each curve is determined by varying all of the effective $J$-factors either up or down through the $1\sigma$ uncertainties.}
  \label{fig:CrossSection}
\end{figure}

To convert the 95\% C.L. bounds on $\PhiPP$ into bounds on particle parameters, we consider dark matter annihilation into four different final states: $b\bar{b}$, $W^+ W^-$, $\mu^- \mu^+$, and $\tau^- \tau^+$.
We use the numerical tools described in Ref.~\cite{Cirelli:2010xx} to obtain the integrated photon spectra at various masses $m_X$ for these annihilation channels.
The left panel of Fig.~\ref{fig:CrossSection} shows the resulting 95\% C.L. bounds in the $(m_X, (\sigma_A v)_0)$ plane for Sommerfeld-enhanced annihilation for {\it Set 7}, while the right panel shows a comparison of the constraints from {\it Set 7} (solid contours) and {\it Set 5} (dotted contours) for the $b\bar{b}$ (red) and $\tau^+\tau^-$ (blue) final states.
In each case, we show the effect on the variation in the 95\% C.L. limits due to the 1$\sigma$ variation in $J$-factors for each model with thin contours in the corresponding style (the area between the uncertainty contours is shaded in all cases except for the {\it Set 5} contours in the right panel of Fig.~\ref{fig:CrossSection}).
One can see that, compared to {\it Set 5}, the smaller uncertainties in the $J$-factors for the {\it Set 7} dwarfs lead to smaller uncertainty bands on $(\sigma_A v)_0$ (which is also the case for $\PhiPP^\bound (\beta)$ in the right panel of Fig.~\ref{fig:N_PhiPPbound}).
Note that it is a coincidence that the upper uncertainty contours for {\it Set 5} and {\it Set 7} overlap and that the central contour for {\it Set 5} is located approximately on top of the lower limit contour for {\it Set 7}.
Additionally, it is clear that the 95\% C.L. bounds are stronger from {\it Set 5} than from {\it Set 7}, though from the right panel of Fig.~\ref{fig:N_PhiPPbound}, this is not the case for all values of $\beta$.

\section{Conclusions and Outlook}
\label{sec:conclusions}

We have introduced \texttt{MADHAT}, an efficient numerical tool that provides statistical limits on the number of observed photons coming from dark matter annihilation/decay or other nonstandard or unknown astrophysics.
\texttt{MADHAT} computes the resulting limits on dark matter annihilation in dwarf satellite galaxies and dwarf candidates for any choice of the dark matter microphysics, astrophysics, or targets that the user makes.
\texttt{MADHAT} is an implementation of the analysis framework presented in Ref.~\cite{Boddy:2018qur}, updated to include the most recent Pass8R3 Fermi-LAT data~\cite{Bruel:2018lac}.
As an example application, we have performed a new analysis with \texttt{MADHAT} to determine bounds on Sommerfeld-enhanced dark matter annihilation for the set of 8 dwarfs considered in Ref.~\cite{Petac:2018gue}.

This tool will be maintained and upgraded as new Fermi-LAT data is released and as new dwarfs are discovered.
In particular, we will update the PMFs and the number of photons observed over the Fermi-LAT exposure.
Although this tool currently utilizes only data from the Fermi-LAT, a similar analysis can be performed using any instrument for which the data is publicly available or for which the appropriate exposures, PMFs, and observed counts are provided by the collaboration.
We anticipate updating this tool to include other instruments in the future.

Finally, we note that \texttt{MADHAT} can be operated as a stand-alone tool, as described here, or incorporated into existing dark matter analysis packages or pipelines.
We are currently working to incorporate \texttt{MADHAT} into the GAMBIT global fitting code for Beyond the Standard Model physics~\cite{Workgroup:2017lvb} so that information from Fermi-LAT dwarf observations can be included in GAMBIT analyses.
We encourage \texttt{MADHAT} to be used in conjunction with other software packages for dark matter analysis as well.

Dwarf satellite galaxies are extremely promising targets for indirect dark matter searches.
Our aim with \texttt{MADHAT} is to enhance the ability of the dark matter community to use indirect detection strategies to learn about the physics of the dark universe.

\acknowledgments
We are grateful to Kyle Kaiser, Savvas Koushiappas, Jack Runburg, and Pat Scott for useful discussions. K.~B., J.~K., and B.~S.~also thank the University of Utah and the \textit{No Stone Unturned} workshop, where parts of this work were completed.
The work of J.~K.~is supported in part by DOE grant DE-SC0010504.
The work of P.~S.~and B.~S.~is supported in part by NSF grant PHY-1720282.

\newpage
\begin{sidewaystable}
  \begin{tabular}{l|l|c|c|c|cccccccccc}
    \hline
    \hline
    \#  & Name & $\Abareff T_\obs$ & $\overline N_\bgd$ & $N_\obs$ & \multicolumn{10}{c}{$\log_{10}(J/[\GeV^2/\cm^5])$} \\
    & & $[10^{11}~\cm^2\s]$ & & & Set 1 & a & b & c & Set 2 & Set 3 & Set 4 & Set 5 & Set 6 & Set 7 \\
    \hline
     1 & Aquarius II       & 4.279 &  139 &  184 & -                       & -                       & -                       & -                       & -                       & -                       & -                       & -                       & $18.27^{+0.66}_{-0.58}$ & -                       \\
     2 & Bootes I          & 4.552 &  143 &  131 & $18.2^{+0.4}_{-0.4}$    & \checkmark              & \checkmark              & \checkmark              & $16.65^{+0.64}_{-0.38}$ & $16.95^{+0.53}_{-0.40}$ & -                       & -                       & $18.17^{+0.31}_{-0.29}$ & -                       \\
     3 & Bootes II         & 4.511 &  143 &  154 & $18.9^{+0.6}_{-0.6}$    & \checkmark              & \checkmark              & -                       & -                       & -                       & -                       & -                       & -                       & -                       \\
     4 & Bootes III        & 4.795 &  122 &  106 & $18.8^{+0.6}_{-0.6}$    & \checkmark              & \checkmark              & \checkmark              & -                       & -                       & -                       & -                       & -                       & -                       \\
     5 & Canes Venatici I  & 4.869 &  108 &   70 & $17.4^{+0.3}_{-0.3}$    & \checkmark              & \checkmark              & \checkmark              & $17.27^{+0.11}_{-0.11}$ & $16.92^{+0.43}_{-0.26}$ & -                       & -                       & $17.42^{+0.17}_{-0.15}$ & -                       \\
     6 & Canes Venatici II & 4.864 &  109 &   92 & $17.6^{+0.4}_{-0.4}$    & \checkmark              & \checkmark              & -                       & $17.65^{+0.40}_{-0.40}$ & $17.23^{+0.84}_{-0.68}$ & -                       & -                       & $17.82^{+0.47}_{-0.47}$ & -                       \\
     7 & Canis Major       & 4.588 &  594 &  431 & -                       & -                       & -                       & -                       & -                       & -                       & -                       & -                       & -                       & -                       \\
     8 & Carina            & 5.055 &  231 &  176 & $17.9^{+0.1}_{-0.1}$    & \checkmark              & \checkmark              & -                       & $17.99^{+0.34}_{-0.34}$ & $17.98^{+0.46}_{-0.28}$ & -                       & -                       & $17.83^{+0.10}_{-0.09}$ & $20.76^{+0.09}_{-0.18}$ \\
     9 & Carina II         & 5.502 &  378 &  346 & -                       & -                       & -                       & -                       & -                       & -                       & -                       & -                       & $18.25^{+0.55}_{-0.54}$ & -                       \\
    10 & Carina III        & 6.029 &  223 &  191 & -                       & -                       & -                       & -                       & -                       & -                       & -                       & -                       & -                       & -                       \\
    11 & Cetus II          & 4.260 &   90 &  107 & $19.1^{+0.6}_{-0.6}$    & -                       & -                       & \checkmark              & -                       & -                       & -                       & -                       & -                       & -                       \\
    12 & Cetus III         & 4.211 &  103 &   83 & -                       & -                       & -                       & -                       & -                       & -                       & -                       & -                       & -                       & -                       \\
    13 & Columba I         & 4.596 &  131 &  133 & $17.6^{+0.6}_{-0.6}$    & -                       & \checkmark              & -                       & -                       & -                       & -                       & -                       & -                       & -                       \\
    14 & Coma Berenices    & 4.587 &  118 &  151 & $19.0^{+0.4}_{-0.4}$    & \checkmark              & \checkmark              & -                       & $18.67^{+0.33}_{-0.32}$ & $18.52^{+0.94}_{-0.74}$ & $18.70^{+0.72}_{-0.69}$ & $21.59^{+0.26}_{-0.29}$ & $19.00^{+0.36}_{-0.35}$ & -                       \\
    15 & Crater II         & 4.285 &  168 &  151 & -                       & -                       & -                       & -                       & -                       & -                       & -                       & -                       & -                       & -                       \\
    16 & Draco             & 6.178 &  196 &  169 & $18.8^{+0.1}_{-0.1}$    & \checkmark              & \checkmark              & -                       & $18.86^{+0.24}_{-0.24}$ & $19.09^{+0.39}_{-0.36}$ & $18.74^{+0.17}_{-0.16}$ & $21.52^{+0.26}_{-0.29}$ & $18.83^{+0.12}_{-0.12}$ & $21.51^{+0.15}_{-0.12}$ \\
    17 & Draco II          & 6.469 &  170 &  167 & $19.3^{+0.6}_{-0.6}$    & \checkmark              & \checkmark              & \checkmark              & -                       & $15.54^{+3.10}_{-4.07}$ & $18.87^{+0.17}_{-0.15}$ & -                       & $18.93^{+1.39}_{-1.70}$ & -                       \\
    18 & Eridanus II       & 4.777 &  105 &   80 & $17.1^{+0.6}_{-0.6}$    & -                       & \checkmark              & \checkmark              & -                       & -                       & -                       & -                       & -                       & -                       \\
    19 & Eridanus III      & 4.911 &  119 &  126 & $18.1^{+0.6}_{-0.6}$    & -                       & -                       & \checkmark              & -                       & -                       & -                       & -                       & -                       & -                       \\
    20 & Fornax            & 4.564 &   98 &  137 & $17.8^{+0.1}_{-0.1}$    & \checkmark              & \checkmark              & \checkmark              & $18.15^{+0.16}_{-0.16}$ & $17.90^{+0.28}_{-0.16}$ & -                       & -                       & $18.09^{+0.10}_{-0.10}$ & $20.99^{+0.39}_{-0.24}$ \\
    21 & Grus I            & 4.900 &  118 &  114 & $17.9^{+0.6}_{-0.6}$    & -                       & \checkmark              & -                       & $17.96^{+0.90}_{-1.93}$ & -                       & -                       & -                       & $16.88^{+1.51}_{-1.66}$ & -                       \\
    22 & Grus II           & 4.670 &  159 &  165 & $18.7^{+0.6}_{-0.6}$    & -                       & \checkmark              & -                       & -                       & -                       & -                       & -                       & -                       & -                       \\
    23 & Hercules          & 4.823 &  251 &  247 & $16.9^{+0.7}_{-0.7}$    & \checkmark              & \checkmark              & \checkmark              & $16.83^{+0.45}_{-0.45}$ & $16.28^{+0.66}_{-0.57}$ & -                       & -                       & $17.37^{+0.53}_{-0.53}$ & -                       \\
    24 & Horologium I      & 5.003 &  121 &  176 & $18.2^{+0.6}_{-0.6}$    & \checkmark              & \checkmark              & -                       & $18.64^{+0.95}_{-0.39}$ & -                       & -                       & -                       & $19.27^{+0.77}_{-0.71}$ & -                       \\
    25 & Horologium II     & 4.899 &  111 &  117 & $18.3^{+0.6}_{-0.6}$    & -                       & \checkmark              & -                       & -                       & -                       & -                       & -                       & -                       & -                       \\
    26 & Hydra II          & 4.577 &  228 &  185 & $17.8^{+0.6}_{-0.6}$    & \checkmark              & \checkmark              & \checkmark              & $16.56^{+0.87}_{-1.85}$ & $13.26^{+2.12}_{-2.31}$ & -                       & -                       & -                       & -                       \\
    27 & Hydrus I          & 5.045 &  217 &  293 & -                       & -                       & -                       & -                       & -                       & -                       & -                       & -                       & -                       & -                       \\
    28 & Indus II          & 4.792 &  239 &  284 & $17.4^{+0.6}_{-0.6}$    & -                       & \checkmark              & \checkmark              & -                       & -                       & -                       & -                       & -                       & -                       \\
    29 & Kim 2             & 4.885 &  220 &  219 & $18.1^{+0.6}_{-0.6}$    & -                       & -                       & \checkmark              & -                       & -                       & -                       & -                       & -                       & -                       \\
    \hline
  \end{tabular}
  \caption{A list of dwarfs (\#1-\#29) used by \texttt{MADHAT}, with parameters described in the text.
  The predefined sets of $J$-factors are {\it Set 1} (and subsets {\it 1a, 1b, 1c})~\cite{Geringer-Sameth:2014yza,Simon:2015fdw}, {\it Set 2}~\cite{Evans:2016xwx}, {\it Set 3}~\cite{Hayashi:2016kcy}, {\it Set 4}~\cite{Ichikawa:2016nbi,Ichikawa:2017rph}, {\it Set 5}~\cite{Boddy:2017vpe}, {\it Set 6}~\cite{Pace:2018tin}, and {\it Set 7}~\cite{Petac:2018gue}.}
  \label{Tab:DsphParametersA}
\end{sidewaystable}

\newpage
\begin{sidewaystable}
  \begin{tabular}{l|l|c|c|c|cccccccccc}
    \hline
    \hline
    \#  & Name & $\Abareff T_\obs$ & $\overline N_\bgd$ & $N_\obs$ & \multicolumn{10}{c}{$\log_{10}(J/[\GeV^2/\cm^5])$} \\
    & & $[10^{11}~\cm^2\s]$ & & & Set 1 & a & b & c & Set 2 & Set 3 & Set 4 & Set 5 & Set 6 & Set 7 \\
    \hline
    30 & Laevens 3         & 4.601 &  248 &  285 & -                       & -                       & -                       & -                       & -                       & -                       & -                       & -                       & -                       & -                       \\
    31 & Leo I             & 4.434 &  126 &  131 & $17.8^{+0.2}_{-0.2}$    & \checkmark              & \checkmark              & \checkmark              & $17.80^{+0.28}_{-0.28}$ & $17.45^{+0.43}_{-0.23}$ & -                       & -                       & $17.64^{+0.14}_{-0.12}$ & $20.57^{+0.16}_{-0.10}$ \\
    32 & Leo II            & 4.497 &  112 &   85 & $18.0^{+0.2}_{-0.2}$    & \checkmark              & \checkmark              & \checkmark              & $17.44^{+0.25}_{-0.25}$ & $17.51^{+0.34}_{-0.28}$ & -                       & -                       & $17.76^{+0.22}_{-0.18}$ & $20.58^{+0.30}_{-0.17}$ \\
    33 & Leo IV            & 4.124 &  132 &  134 & $16.3^{+1.4}_{-1.4}$    & \checkmark              & \checkmark              & -                       & $16.64^{+0.90}_{-0.90}$ & $15.31^{+1.58}_{-2.90}$ & -                       & -                       & $16.40^{+1.01}_{-1.15}$ & -                       \\
    34 & Leo T             & 4.434 &  129 &  120 & -                       & -                       & -                       & -                       & $17.32^{+0.38}_{-0.37}$ & $16.75^{+0.61}_{-0.53}$ & -                       & -                       & $17.49^{+0.49}_{-0.45}$ & -                       \\
    35 & Leo V             & 4.139 &  130 &  152 & $16.4^{+0.9}_{-0.9}$    & \checkmark              & \checkmark              & \checkmark              & $16.94^{+1.05}_{-0.72}$ & $16.24^{+1.26}_{-1.36}$ & -                       & -                       & $17.65^{+0.91}_{-1.03}$ & -                       \\
    36 & Pegasus III       & 4.243 &  168 &  182 & $17.5^{+0.6}_{-0.6}$    & -                       & \checkmark              & \checkmark              & -                       & -                       & -                       & -                       & $18.30^{+0.89}_{-0.97}$ & -                       \\
    37 & Phoenix II        & 4.853 &  117 &   98 & $18.1^{+0.6}_{-0.6}$    & -                       & \checkmark              & \checkmark              & -                       & -                       & -                       & -                       & -                       & -                       \\
    38 & Pictor I          & 4.958 &  123 &  118 & $17.9^{+0.6}_{-0.6}$    & -                       & \checkmark              & \checkmark              & -                       & -                       & -                       & -                       & -                       & -                       \\
    39 & Pictor II         & 5.365 &  263 &  307 & -                       & -                       & -                       & -                       & -                       & -                       & -                       & -                       & -                       & -                       \\
    40 & Pisces II         & 4.224 &  160 &  139 & $17.6^{+0.6}_{-0.6}$    & \checkmark              & \checkmark              & \checkmark              & $17.90^{+1.14}_{-0.80}$ & $15.94^{+1.25}_{-1.28}$ & -                       & -                       & $17.30^{+1.00}_{-1.09}$ & -                       \\
    41 & Reticulum II      & 5.039 &  119 &  134 & $18.9^{+0.6}_{-0.6}$    & \checkmark              & \checkmark              & \checkmark              & $18.71^{+0.84}_{-0.32}$ & $17.76^{+0.93}_{-0.90}$ & -                       & $21.67^{+0.33}_{-0.30}$ & $18.96^{+0.38}_{-0.37}$ & -                       \\
    42 & Reticulum III     & 5.302 &  139 &  174 & $18.2^{+0.6}_{-0.6}$    & -                       & \checkmark              & \checkmark              & -                       & -                       & -                       & -                       & -                       & -                       \\
    43 & Sagittarius       & 4.760 &  564 &  637 & -                       & -                       & -                       & -                       & -                       & -                       & -                       & -                       & -                       & -                       \\
    44 & Sagittarius II    & 4.640 &  344 &  323 & $18.4^{+0.6}_{-0.6}$    & -                       & \checkmark              & \checkmark              & -                       & -                       & -                       & -                       & -                       & -                       \\
    45 & Sculptor          & 4.435 &   93 &  123 & $18.5^{+0.1}_{-0.1}$    & \checkmark              & \checkmark              & -                       & $18.65^{+0.29}_{-0.29}$ & $18.42^{+0.35}_{-0.17}$ & -                       & -                       & $18.58^{+0.05}_{-0.05}$ & $21.37^{+0.16}_{-0.13}$ \\
    46 & Segue 1           & 4.399 &  126 &  150 & $19.4^{+0.3}_{-0.3}$    & \checkmark              & \checkmark              & \checkmark              & $19.41^{+0.39}_{-0.40}$ & $17.95^{+0.90}_{-0.98}$ & $19.81^{+0.93}_{-0.74}$ & $22.25^{+0.37}_{-0.62}$ & $19.12^{+0.49}_{-0.58}$ & -                       \\
    47 & Segue 2           & 4.556 &  225 &  268 & -                       & -                       & -                       & -                       & $17.11^{+0.85}_{-1.76}$ & $13.09^{+1.85}_{-2.62}$ & -                       & -                       & -                       & -                       \\
    48 & Sextans           & 4.145 &  133 &  152 & $17.5^{+0.2}_{-0.2}$    & \checkmark              & \checkmark              & -                       & $17.87^{+0.29}_{-0.29}$ & $17.71^{+0.39}_{-0.21}$ & -                       & -                       & $17.73^{+0.13}_{-0.12}$ & $20.54^{+1.37}_{-0.34}$ \\
    49 & Triangulum II     & 4.876 &  203 &  216 & $19.1^{+0.6}_{-0.6}$    & \checkmark              & \checkmark              & -                       & -                       & $20.44^{+1.20}_{-1.17}$ & -                       & -                       & -                       & -                       \\
    50 & Tucana II         & 5.085 &  133 &  135 & $18.6^{+0.6}_{-0.6}$    & \checkmark              & \checkmark              & -                       & $19.05^{+0.87}_{-0.58}$ & -                       & -                       & -                       & $18.84^{+0.55}_{-0.50}$ & -                       \\
    51 & Tucana III        & 5.098 &  121 &  141 & $19.3^{+0.6}_{-0.6}$    & -                       & \checkmark              & \checkmark              & -                       & -                       & -                       & -                       & -                       & -                       \\
    52 & Tucana IV         & 5.148 &  123 &  117 & $18.7^{+0.6}_{-0.6}$    & -                       & \checkmark              & \checkmark              & -                       & -                       & -                       & -                       & -                       & -                       \\
    53 & Tucana V          & 5.204 &  130 &  109 & $18.6^{+0.6}_{-0.6}$    & -                       & -                       & \checkmark              & -                       & -                       & -                       & -                       & -                       & -                       \\
    54 & Ursa Major I      & 5.519 &  118 &  114 & $17.9^{+0.5}_{-0.5}$    & \checkmark              & \checkmark              & -                       & $18.48^{+0.25}_{-0.25}$ & $17.48^{+0.42}_{-0.30}$ & $18.67^{+1.75}_{-1.02}$ & -                       & $18.26^{+0.29}_{-0.27}$ & -                       \\
    55 & Ursa Major II     & 6.315 &  202 &  254 & $19.4^{+0.4}_{-0.4}$    & \checkmark              & \checkmark              & -                       & $19.38^{+0.39}_{-0.39}$ & $19.56^{+1.19}_{-1.25}$ & $19.50^{+0.29}_{-0.30}$ & -                       & $19.44^{+0.41}_{-0.39}$ & -                       \\
    56 & Ursa Minor        & 6.828 &  163 &  142 & $18.9^{+0.2}_{-0.2}$    & \checkmark              & \checkmark              & -                       & $19.15^{+0.25}_{-0.24}$ & -                       & $19.12^{+0.15}_{-0.12}$ & $21.69^{+0.27}_{-0.34}$ & $18.75^{+0.12}_{-0.12}$ & $21.63^{+0.21}_{-0.13}$ \\
    57 & Virgo I           & 4.139 &  130 &  132 & -                       & -                       & -                       & -                       & -                       & -                       & -                       & -                       & -                       & -                       \\
    58 & Willman 1         & 5.414 &  116 &  133 & $18.9^{+0.6}_{-0.6}$    & \checkmark              & \checkmark              & \checkmark              & $19.29^{+0.91}_{-0.62}$ & -                       & -                       & -                       & $19.53^{+0.50}_{-0.50}$ & -                       \\
    \hline
  \end{tabular}
  \caption{A list of dwarfs (\#30-\#58) used by \texttt{MADHAT}, with parameters described in the text.
  The predefined sets of $J$-factors are {\it Set 1} (and subsets {\it 1a, 1b, 1c})~\cite{Geringer-Sameth:2014yza,Simon:2015fdw}, {\it Set 2}~\cite{Evans:2016xwx}, {\it Set 3}~\cite{Hayashi:2016kcy}, {\it Set 4}~\cite{Ichikawa:2016nbi,Ichikawa:2017rph}, {\it Set 5}~\cite{Boddy:2017vpe}, {\it Set 6}~\cite{Pace:2018tin}, and {\it Set 7}~\cite{Petac:2018gue}.}
  \label{Tab:DsphParametersB}
\end{sidewaystable}


\end{document}